\documentclass[letterpaper, 10 pt, conference]{ieeeconf}

\IEEEoverridecommandlockouts
\usepackage{amsmath}
\usepackage{amssymb}
\usepackage{xfrac}
\usepackage{braket}
\usepackage{qcircuit}
\graphicspath{ {./images/} }

\newcommand{\lft}{\text{\textbf{left}}}
\newcommand{\rht}{\text{\textbf{right}}}
\newcommand{\up}{\text{\textbf{up}}}
\newcommand{\dwn}{\text{\textbf{down}}}

\def\nand{\tilde\land}

\title{\LARGE \bf
Efficient circuit implementation for coined quantum walks on binary trees and application to reinforcement learning
}

\author{Thomas Mullor\\IRT Saint Exupery\\Toulouse\\thomas.mullor@irt-saintexupery.com \and David Vigouroux \\ IRT Saint Exupery\\Toulouse\\david.vigouroux@irt-saintexupery.com \and Louis Bethune \\ Université Paul Sabatier, IRIT\\Toulouse\\louis.bethune@univ-toulouse.fr}

\begin{document}

\maketitle
\thispagestyle{empty}
\pagestyle{empty}

\begin{abstract}

Quantum walks on binary trees are used in many quantum algorithms to achieve important speedup over classical algorithms. The formulation of this kind of algorithms as quantum circuit presents the advantage of being easily readable, executable on circuit based quantum computers and simulators and optimal on the usage of resources. We propose a strategy to compose quantum circuit that performs quantum walk on binary trees following universal gate model quantum computation principles. We give a particular attention to NAND formula evaluation algorithm as it could have many applications in game theory and reinforcement learning. We therefore propose an application of this algorithm and show how it can be used to train a quantum reinforcement learning agent in a two player game environment.

\end{abstract}

\section{INTRODUCTION}
Quantum computing is a computation paradigm using properties of quantum mechanics to perform information processing. Many famous quantum algorithms have been shown to outperform their equivalent classical algorithm\cite{grover, shor}. Quantum walk is a way to compose many promising quantum algorithms. It can be viewed as the quantum analogues of classical random walks \cite{quantum_walk_intro}. In several studies, it has been shown that it could provide some algorithmic speedup on many problems \cite{speedup_quantum_walk, element_distinctness, triangle_finding} and possible applications to tree search problems \cite{search_via_quantum_walk, any_and_or}.\\

To consider a quantum algorithm as implementable on universal quantum computers, we describe them as a quantum circuit using native quantum gates. For particular cases of quantum walks, circuit implementation has been proposed allowing us to execute them easily with an optimal usage of resources \cite{efficient_q_circ}. To the best of our knowledge, no circuit based implementation has been presented for quantum walks on binary trees even if existing algorithms uses such processes. In this work, we propose a strategy to perform implementation of such a quantum walk using quantum gate model where the number of gates used for a walk step scales linearly with the depth of the tree.
We will give a particular interest to quantum walk based algorithm performing boolean NAND tree evaluation\cite{any_and_or} as it has potential applications to game tree resolution and reinforcement learning.\\

Reinforcement learning is an area of machine learning where an agent learn to take the best actions in a given environment to maximize a reward. Many classical reinforcement learning algorithms showed impressive results in two player games like chess or go \cite{mastering_chess, mastering_go}. Most of the algorithms who try to master a game relies on tree search algorithms like Monte Carlo Tree Search (MCTS) to explore game tree and choose best move \cite{mastering_chess, mastering_go}. This technique is very powerful but sometimes has flaws as it performs a partial exploration of the game tree. In this article, we are interested in the performances of quantum computing for exploring game tree. Improving tree search algorithm could improve training and results of a reinforcement learning agent.\\

As NAND formula algorithm allow us to evaluate quality of a position in a two-player game tree, we illustrate its potential application by using it as a training tool for a quantum agent in a simple two-player game. 
With the speed-up proposed by this algorithm, we are able to perform evaluation of deeper trees in equivalent time (twice deeper exploration for a binary tree). By using quantum algorithm to perform such explorations, we expect agents to achieve better performances in their learning process.\\

\section{RELATED WORKS}

Many quantum algorithms based on quantum walks have been presented with various application fields. For some kinds of walks on very specifics graphs (cycles, hypercubes, welded trees, highly symmetric graphs...), efficient quantum circuit designs have been presented\cite{efficient_q_circ} and allows us to easily implement quantum walks algorithms using circuit-based quantum computers.

One of the main quantum algorithm based on quantum walks on binary trees allows us to evaluate a boolean formula of \(N\) variables taking the form of a NAND tree using only \(N^{1/2+o(1)}\) calls to a black box oracle\cite{any_and_or}. By extension, this allows us to evaluate any AND/OR tree (i.e. a tree where vertices correspond alternately to AND and OR evaluations), which is the boolean version of a MIN/MAX tree. As MIN/MAX trees represents optimal decision makings in a two-player game, its computation constitutes an useful knowledge to learn to play a game. The work presented in this paper is motivated by the potential applications of this algorithm in game theory and reinforcement learning.

Quantum NAND formula evaluation algorithm is presented as a quantum phase estimation performed on a quantum walk operator using an input oracle. 
The operator performs a quantum walk on a rooted binary tree, i.e. a binary tree on which we added two-vertices tail attached to the root of the tree. The input oracle gives values of the leaves of the tree and is incorporated to the walk by turning the leaves evaluating to 1 to probability sinks.\\

\begin{figure}[h]
\centerline{\includegraphics[width=0.3\textwidth]{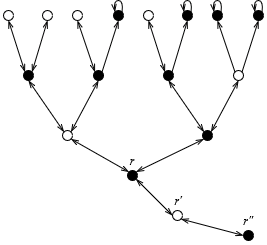}}
\caption{The rooted binary tree for the simple boolean formula $\varphi=\left(\left(x_{1} \nand x_{2}\right) \nand\left(x_{3} \nand x_{4}\right)\right) \nand\left(\left(x_{5} \nand x_{6}\right) \nand\left(x_{7} \nand x_{8}\right)\right)$ where internal vertices is result of NAND evaluation of their children. Vertices are filled if they evaluate to 1. A root made of two vertices ($r'$ and $r''$) has been added to the tree. Leaves evaluating to 1 are turned to probability sinks as presented in \cite{any_and_or}} 
\label{rooted_tree}
\end{figure}

Let us remind from \cite{any_and_or} that evaluating a binary NAND tree allows us to evaluate any AND/OR tree after an efficient classical preprocessing of the formula taking $\text{poly}(N)$ time. \\

\textbf{Contributions} \\

In this paper, we propose an efficient circuit based implementation of quantum walks on a binary tree and we propose, as an application, a full implementation of the main components of NAND formula evaluation algorithm where the number of qubits and gates grows linearly with the depth of the tree.

In a second part, we illustrate the usage of NAND formula evaluation, using it in a reinforcement learning case by proposing a way to learn the input oracle.

\section{Quantum Circuits for coined quantum walk on an arbitrary binary tree}

We can describe a coined quantum walk on a graph \(G = (V, E)\) as a process defined by two operators acting on their associated qubits registers : \\
    - Coin : a flip operator $F$ acts on a coin register $C$. Superposition of states of $C$ will indicate which paths the walker must take.\\
    - Walker : a shift operator $S$ acts on a walker register $W$. Depending on the state of $C$, the state of $W$ will change to go through the different vertices $v$ of $G$.\\
The two operators define a walk step operator $U = SF$. For a given superposition of vertices $v_t$ held in a register \(W\) and an action $a_t$ held in a register \(C\) : 
\[U|v_t\rangle|a_{t}\rangle\ = |v_{t+1}\rangle|a_{t+1}\rangle.\]
To facilitate execution of such a process on a quantum computer, it is preferable that the shift operator can be easily expressed as a quantum circuit of limited size. We therefore propose a vertex labelling that allows us to express shift operator with elementary mathematical operations and show how we can implement quantum algorithm based specifically on walks on binary trees.

\subsection{Labeling of the nodes of the tree and shift operator}
As we store the label of the vertices we go through in the \(W\) qubits register, the way we label the vertices of our tree will have a huge influence on the difficulty to implement the shift operator. Finding a labeling of the vertices that allows us to express easily the transition between any states as a simple operation and thus express easily the shift operator is a key step. We thus propose to label the vertices of the tree by following labelling, which have such property.

We will consider here an arbitrary binary tree $T$ we want to perform quantum walk on. 
We perform labeling of the vertices of the tree $T$ considering the following rules:
\begin{enumerate}
    \item root of $T$ is vertex 1
    \item left child of a vertex $v$ is labeled $2v$
    \item right child of a vertex $v$ is labeled $2v + 1$
\end{enumerate}
For the case of perfectly balanced binary trees, this corresponds to a breadth first labelling and is the canonical labelling. It allows to implement easily all the operations used to build the shift operator with a basic set of quantum gates.
\begin{figure}[h]
\centerline{\includegraphics[width=0.3\textwidth]{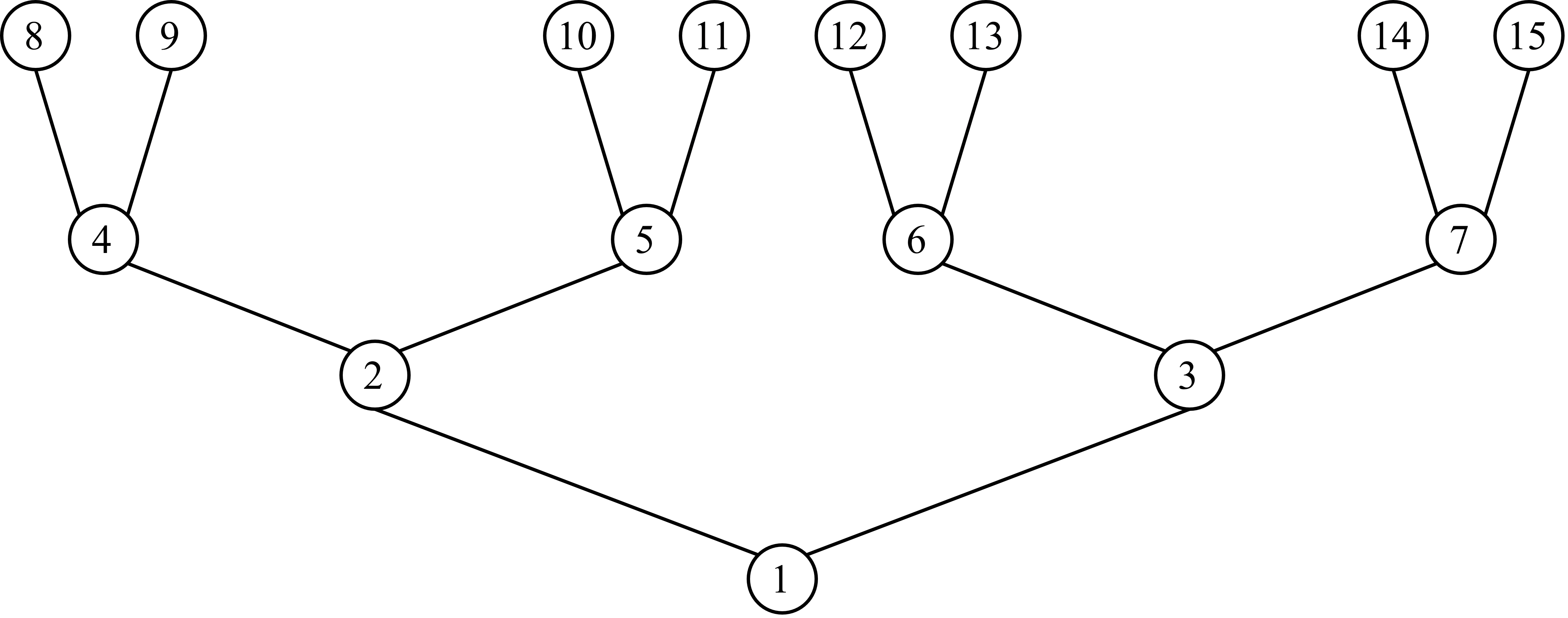}}
\caption{Example of labelling on a balanced binary tree} 
\label{example}
\end{figure}\\
With this labeling, we can define the shift application \(S:|v_t\rangle|a\rangle \mapsto |v_{t+1}\rangle|f(v_t, a_t)\rangle\) allowing us to perform steps through the vertices of the tree starting from any vertex $v$ :

\begin{equation}
    \begin{aligned}
    S|v\rangle|\lft\rangle &= |2v\rangle|f(v, \lft)\rangle\\
    S|v\rangle|\rht\rangle &= |2v + 1\rangle|f(v, \rht)\rangle\\
    S|v\rangle|\up\rangle &= \left\{
    \begin{array}{ll}
        |v/2\rangle|f(v, \up)\rangle & \mbox{if } v \mbox{ is even,} \\
        |(v-1)/2\rangle|f(v, \up)\rangle & \mbox{if } v \mbox{ is odd.}
    \end{array}
    \right.
    \end{aligned}
\end{equation}

For now, we are only interested in the transformation performed on the value held in the register \(W\) by an application of $S$. The transformation performed on the coin \(C\) (here represented by $f(v_t, a_t)$) can be any transformation as long as $S$ is a valid (unitary) application and depends on the algorithm we are implementing. Example of $S$ operator with detailed $f$ will be given further in this paper.\\

The register \(W\) holding the position of the walker requires a number of qubits equals to the maximal depth \(d\) of the tree. The register must be able to store, using basis encoding, the highest possible label of a vertex of the tree which is \(2^{d} - 1\) and can thus be stored in a \(log_2(2^d) = d\) qubits register. Having an adapted register, we can implement the $f:|v\rangle \mapsto |v+1\rangle$ operation used in the left step with the following circuit \(P\) (with the highest qubit as the most significant qubit) :
\[
\Qcircuit @C=1em @R=.9em {
&&\qw & \targ & \qw & \qw & \qw & \qw & \qw & \qw\\
&&\qw & \ctrl{-1} & \targ & \qw & \qw & \qw & \qw & \qw\\
&&\vdots  & & & \hdots\\
P = &\\
&&\qw & \ctrl{-4} & \ctrl{-3} & \qw &  \targ & \qw & \qw & \qw\\
&&\qw & \ctrl{-5} & \ctrl{-4} & \qw & \ctrl{-1} &  \targ & \qw & \qw\\
&&\qw & \ctrl{-6} & \ctrl{-5} & \qw & \ctrl{-1} & \ctrl{-1} & \targ & \qw &\\
}\]
The $f:|v\rangle \mapsto |v-1\rangle$ application used in the down step is implemented with the $P^{\dagger}$ circuit :\\
\[
\Qcircuit @C=1em @R=.9em {
&&\qw & \qw & \qw & \qw & \qw & \qw & \targ & \qw\\
&&\qw & \qw & \qw & \qw & \qw & \targ & \ctrl{-1}  & \qw\\
&&&&&& \hdots & & & \vdots \\
P^{\dagger} = &\\
&&\qw & \qw & \qw & \targ & \qw & \ctrl{-3} & \ctrl{-4} & \qw\\
&&\qw & \qw& \targ & \ctrl{-1} & \qw  & \ctrl{-4} & \ctrl{-5} & \qw \\
&&\qw & \targ & \ctrl{-1} & \ctrl{-1} & \qw & \ctrl{-5} & \ctrl{-6} & \qw  \\
}\]

The multi-controlled-$X$ gates used in this circuit are not native gates but can be implemented with a linear number of elementary gates using the method proposed in \cite{he2017decompositions, decomposition_of_toffoli}.\\

The $f:|v\rangle \mapsto |v*2\rangle$ application used in both left and right steps can be approximately implemented with a circular left shift circuit $M$ made of $SWAP$ gates :

\[
\Qcircuit @C=1.3em @R=1.1em {
&& \qw & \qw & \qw & \qw & \qswap & \qw &&& \qw & \link{1}{-1} & \qw\\
&& \qw & \qw & \qw & \qswap & \qw \qwx & \qw &&& \qw & \link{1}{-1} & \qw\\
&&\vdots  & & \hdots & \qwx & \qwx &&& \vdots &&& \vdots\\
M = &&&&& \qwx & \qwx && = &&&\link{1}{-1}\\
&& \qw & \qswap & \qw & \qw\qwx & \qw\qwx & \qw &&& \qw & \link{1}{-1} & \qw\\
&& \qswap & \qw\qwx & \qw& \qw\qwx & \qw\qwx & \qw &&& \qw & \link{1}{-1} & \qw\\
&& \qswap \qwx & \qswap \qwx & \qw & \qswap \qwx & \qswap \qwx & \qw &&& \qw& \link{-6}{-1} & \qw\\
}\]\\

We can note that this circuit only implements $f:|v\rangle \mapsto |v*2\rangle$ application if the most significant qubit of the \(W\) register is in $|0\rangle$ state, otherwise the least significant qubit (who should always be zero as $2v$ is obviously an even number) would be swapped with a qubit in the $|1\rangle$ state. This inconvenience can be ignored in the case of our quantum walk on a tree as the only situations where the most significant qubit of \(W\) will be in $|1\rangle$ state is the case where the label of the vertex \(v\) stored in \(W\) is greater than \(2^{d-1}\). This corresponds to the deepest vertices (i.e. leaves) of the tree which have no child and thus are not eligible nodes for $f:|v\rangle \mapsto |v*2\rangle$ operation as it is used to perform steps that goes deeper in the tree.
The $M$ circuit operator therefore presents all the properties required to implement the transformation allowing us to perform a step on the tree.\\

The $f:|v\rangle \mapsto |v/2\rangle$ application used in down steps can be approximately implemented with the $M^{\dagger}$ circular right shift circuit. This circuit only implements $|v\rangle \mapsto |v/2\rangle$ if $v$ is an even number which is guaranteed by the labeling method and shift operator properties we defined.\\

A quantum walk on a binary tree also needs a coin register \(C\) which is a qutrit holding three states : $|\dwn\rangle$, $|\lft\rangle$ or $|\rht\rangle$. For the convenience of the implementation, we simply materialize it by a 2 qubits register and use the arbitrarily chosen following states :

\begin{equation}
    |\dwn\rangle = |00\rangle\qquad |\lft\rangle = |10\rangle\qquad |\rht\rangle = |11\rangle
\end{equation}
    
With this representation, we can implement a whole step of coined quantum walk using calls to the different operators defined above controlled by the qubits of the coin register.

\subsection{Implementation of A. Childs \& Al boolean formula evaluation algorithm}

NAND tree evaluation performs a quantum phase estimation on a quantum walk operator. The quantum walk is performed on a binary tree extended with a root of two vertices (\(r'\) and \(r''\)) and where leaves evaluating to 1 are turned to probability sinks.

The added root does not change our labeling method in any way : new root \(r''\) will now be labeled at 0, \(r'\) at 1  and other vertices following rules defined previously starting from 2 (new \(r\) value). This forces us to increase the size of the \(W\) register by 1 qubit as we increase the total depth of the tree. 

The quantum walk operator is defined in two parts : the diffusion step and the walk step. The walk step, is a simple application that transforms the state of \(W\) to make it perform a step in the direction specified by the coin and change the value of the coin to make it point the vertex where the walker comes from.

Considering this definition and the labeling we propose, the walk step operator $U$ is thereby defined by an unitary transformation having the following properties :
\begin{equation}
    \begin{aligned}
    U|2*k\rangle|\dwn\rangle &= |k\rangle|\lft\rangle\\
    U|2*k+1\rangle|\dwn\rangle &= |k\rangle|\rht\rangle\\
    U|k\rangle|\lft\rangle &= |2*k\rangle|\dwn\rangle\\
    U|k\rangle|\rht\rangle &= |2*k+1\rangle|\dwn\rangle\\
    \end{aligned}
\end{equation}

As we can write this transformation with the operations we have found circuits for and simple modification of the coin, we can implement this walk step by simply making controlled calls to previously defined circuits. We also have to assure that the coin register ends up in the appropriate state. This gives the circuit presented in Fig. 3 which exactly describes the quantum walk step described above on a binary tree where vertices are labeled as proposed. The fully detailed gate description of this circuit can be found in appendix A.\\

\begin{figure}[h]
\hspace{0.5 cm}
\resizebox{220pt}{\height}{ 
\Qcircuit @C=1.3em @R=1.1em {
    & \multigate{5}{M} & \multigate{5}{P} & \qw & \qw & \multigate{5}{P^{\dagger}} & \multigate{5}{M^{\dagger}} & \qw & \qw & \qw & \qw\\
    & \ghost{M} & \ghost{P} & \qw & \qw & \ghost{P^{\dagger}} & \ghost{M^{\dagger}} & \qw & \qw & \qw & \qw\\
    \vdots &&& \vdots &&&&& \vdots\\
    &\\
    & \ghost{M} & \ghost{P} & \qw & \qw & \ghost{P^{\dagger}} & \ghost{M^{\dagger}} & \qw & \qw & \qw & \qw\\
    & \ghost{M} & \ghost{P} & \qw & \ctrl{2}  & \ghost{P^{\dagger}} & \ghost{M^{\dagger}} & \qw & \ctrl{2} & \qw & \qw \inputgroupv{1}{6}{1em}{4em}{W}\\
    & \ctrl{-1} & \ctrl{-1} & \targ & \ctrl{1} & \ctrl{-1}& \ctrl{-1} & \targ & \ctrl{1} & \targ & \qw\\
    & \qw & \ctrl{-2} &\qw & \targ & \ctrl{-2} & \qw & \qw & \targ & \qw & \qw \inputgroupv{7}{8}{1em}{1em}{C\,}
}
}
\caption{Walk step circuit}
\end{figure}
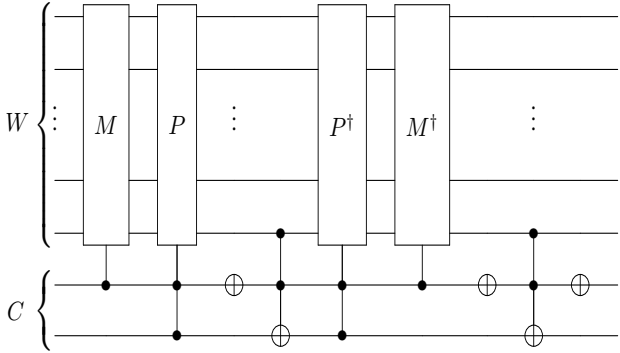 

The diffusion step is described in \cite{any_and_or} as an application of the oracle if we are on a leaf, an application of operator $\text{Reflection}_{|u\rangle}$ if we are in an internal vertice and an application of $\text{Reflection}_{|u'\rangle}$ if we are in $r' = 1$. For a perfectly balanced tree, as the most significant qubit of $W$ indicates whether we are on a leaf or not, the diffusion circuit can be easily defined as presented in Fig. 4 with $O_f$ the input oracle and $R_{|u\rangle}$ and $R_{|u'\rangle}$ the two diffusion operators proposed in \cite{any_and_or}.

\begin{figure}[h]
\hspace{0.5 cm}
\resizebox{220pt}{\height}{ 
\Qcircuit @C=1.3em @R=1.1em {
    & \multigate{5}{O_f} & \targ & \ctrl{6} & \ctrl{1} & \qw & \ctrl{1} & \ctrl{1} & \targ & \qw \\
    & \ghost{O_f} & \targ & \qw & \ctrl{3} & \qw & \ctrl{3} & \ctrl{3} & \targ & \qw \\
    \vdots && \vdots &&&&&& \vdots\\
    &\\
    & \ghost{O_f} & \targ & \qw & \ctrl{1} & \qw & \ctrl{1} & \ctrl{1} & \targ & \qw \\
    & \ghost{O_f} & \targ & \qw & \ctrl{1} & \targ & \ctrl{1} & \ctrl{1} & \qw & \qw \inputgroupv{1}{6}{1em}{4em}{W}\\
    & \qw & \qw & \multigate{1}{R_{|u\rangle}} & \multigate{1}{R_{|u\rangle}} & \qw & \multigate{1}{R_{|u\rangle}} & \multigate{1}{R_{|u'\rangle}} & \qw & \qw \\
    & \qw & \qw & \ghost{R_{|u\rangle}} & \ghost{R_{|u\rangle}} & \qw & \ghost{R_{|u\rangle}} & \ghost{R_{|u'\rangle}} & \qw & \qw \inputgroupv{7}{8}{1em}{1em}{C\,}%
}
}
\caption{Diffusion step circuit}
\end{figure}
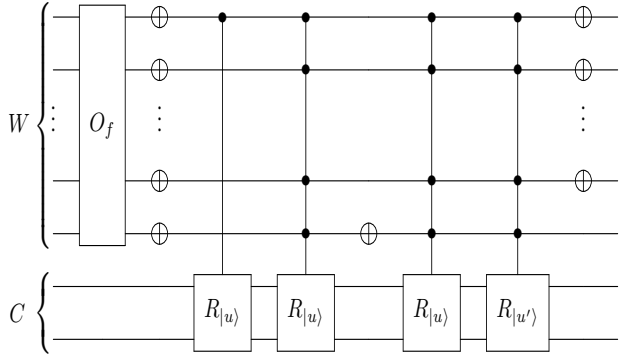 

$R_{|u'\rangle}$ is defined in \cite{any_and_or} as:

\begin{equation}
    \begin{aligned}
    R_{|u'\rangle} &= 2 * |u'\rangle\langle u'| - I\\
    \text{ with }|u'\rangle &= \frac{1}{\sqrt[\leftroot{2} \uproot{2} 4]{N}}|\dwn\rangle + \sqrt{1 - \frac{1}{\sqrt(N)}}|\lft\rangle.
    \end{aligned}
\end{equation}

This can be exactly implemented with the circuit:

\[
\Qcircuit @C=1em @R=.9em {
& \ctrl{1} & \gate{Z} & \qw & \gate{RY(\theta)} & \qw & \qw\\
& \gate{Z} & \gate{Z} & \gate{X} & \ctrl{-1} & \gate{X} & \qw
}\]

with
\[\theta = 2\left(2\pi - cos^{-1}\left(\frac{2-\sqrt(N)}{\sqrt(N)}\right)\right).\]

And $R_{|u\rangle}$ is defined as :
\begin{equation}
    \begin{aligned}
    R_{|u\rangle} &= 2 * |u\rangle\langle u| - I\\
    \text{ with }|u\rangle &= \frac{1}{\sqrt{3}}(|\dwn\rangle + |\lft\rangle + |\rht\rangle).
    \end{aligned}
\end{equation}

This unitary transformation can be implemented with the following circuit : \\

\begin{figure}[h]
\resizebox{230pt}{\height}{ 

\Qcircuit @C=1em @R=.9em {
& \qw & \gate{RY(\frac{\pi}{2})} & \qw & \gate{Z} & \qw & \gate{RY(-\frac{\pi}{2})} & \qw & \qw\\
& \gate{X} & \ctrl{-1} & \gate{RY(\theta_1)} & \ctrl{-1} & \gate{RY(\theta_2)} & \ctrl{-1} & \gate{X} & \qw\\
}
}
\end{figure}

with :
\[\theta_1 =-\cos^{-1}(1/3).\]
\[\theta_2 = \cos^{-1}(-1/3) + \pi.\]

Using this implementation, the number of qubits and the number of gates required grows in $O(d)$ which, for a balanced binary tree is $O(log(N))$. 

All the circuits presented in this section have been implemented and tested using MyQLM quantum simulators. 

For the full implementation of this algorithm, we need an oracle, a transformation $O_f:|x\rangle\mapsto(-1)^{f(x)}|x\rangle$ for a given boolean function $f : \{0, 1\}^n \rightarrow \{0, 1\}$ giving us values of the leaves of the tree.\\

It has been shown that an adapted tunable quantum circuit could be parameterized to reproduce any boolean formula \cite{tunable_qnn_bool_f}. Such a circuit could make an oracle adapted to any NAND formula. As this circuit has to be applied only if we are on a leaf and we want the result of the evaluated function as a phase flip on the \(W\) register, we would have to replace the $X$ transformations used on the result qubit in \cite{tunable_qnn_bool_f} by $Z$ transformations. For cases of perfectly balanced binary trees, this $Z$ gate would be applied on the most significant qubit of the $W$ register as the value of this qubit indicates whether we are on a leaf or not. The second most significant qubit of the $W$ register should never be used to control the $Z$ transformation as, with the labelling we defined, no leave of the tree would have a label where the two most significant qubits are in the $|1\rangle$ state.\\

We will show in the next section how such a circuit can be trained to learn oracle knowledge by evaluating and improving the ability of the circuit to fit a real tree in a two player game.\\

\section{application of parameterized NAND-formula evaluation algorithm to learn a 2 player game by predicting oracle}

In a two-player game, the decision tree of a game in which each player has full and perfect knowledge about the environment can be viewed as a MIN/MAX tree, a tree describing the process where the first player tries to maximize his chance to win the game and second player tries to minimize it. For a simple win-or-lose game, this tree takes the form of a boolean formula tree alternating AND (boolean min) and OR (boolean max) at each node and where the leaves evaluates to 1 if first player wins and 0 otherwise.

We introduced in previous section the possibility to replace the oracle by a tunable quantum circuit able to fit any AND/OR formula. With a circuit of this type, finding the appropriate oracle for a given problem correspond to find the weights of the circuit that make it fit the wanted boolean function. In other terms, this is a combinatorial optimisation problem where we search the combination of activation of gates that gives us the good oracle.\\

To solve this problem, we need to store the labels of the vertices of the NAND tree describing the game in the \(W\) register. This represents even for simple games an amount of qubits we are not in position to get or to simulate. We therefore will apply this algorithm to solve subtrees of a game. We define a game tree of limited size, {\it locally} valid at a given state and for a given horizon. To face the huge amount of subtree we would have to deal with and therefore the huge amount of oracle we would have to find, these oracles will be predicted by a neural network and the exploration of action space will be done by resolving the game tree by the quantum algorithm previously described. \\

\begin{figure}[h]
\centerline{\includegraphics[width=0.45\textwidth]{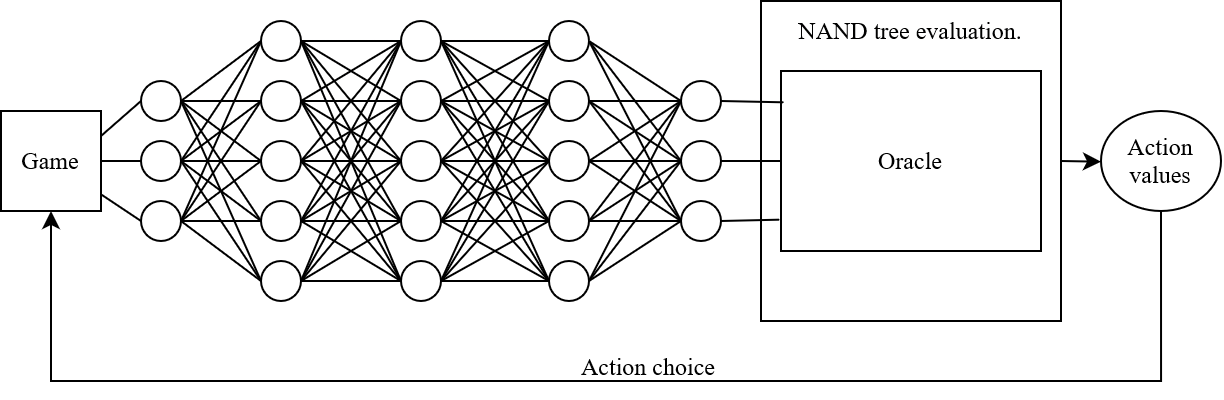}}
\caption{A neural network takes as input information about current state of the game and predict adequate oracle for subtree. NAND tree evaluation is made for this subtree and action is taken by agent according to result of NAND tree evaluation} 
\label{5}
\end{figure}

We give in appendix B more details about the policy used by the agent to take decision in game using NAND formula evaluation.\\

{\bf Game Tree parametrization and neural network}

To implement this method, the neural network has to predict a quantum circuit who will take the place of the oracle in NAND formula evaluation. We propose two strategies to allow the neural network to make such a prediction.

The first strategy is to prepare the oracle as a fully expressive tunable parameterized quantum circuit for boolean function as presented in \cite{tunable_qnn_bool_f} with slight modification explained in the previous section.
 The neural network will thus have to predict the weights of this circuit (which are binary values). This method allows the parameterized quantum circuit to take any boolean formula and thus gives us the guarantee that any oracle can be done to face any state of the game. Nevertheless, it requires a large amount of quantum gates.\\

The second strategy allows us to reduce the number of required gates by forcing the circuit to be of limited depth. Instead of predicting activation of the gates of a fixed structure parameterized quantum circuit, we define a maximal number of gates and train the neural network to predict directly the gates who will compose the circuit : this can be implemented by making the neural network predicting the matrix corresponding to the positions of control of each multi-controlled $Z$ gate applied on the most significant qubit of the $W$ register (already controlled by second most significant qubit). This method allow us to limit the number of gates composing the oracle circuit but may not be able to express any boolean function depending on the maximum number of gates the circuit can be made of. This is the method that we selected for our strategy which is a good balance between resources usage and performance.\\

Consequently, the neural network will be a feed forward neural network which computes the matrix of multiple controls of $Z$ gates composing the oracle circuit for each possible action at a given state of the game. \\

{\bf Learning Procedure}

Our method presents the inconveniences that our predictor is not differentiable. The neural network will thus be trained in a simple reinforcement learning process with a genetic algorithm \cite{deep_neuroevolution} : a wide population of predictors will be initialized and fitness of agents will be calculated by letting them play a two-player game. Best predictors will be selected, mutated and crossed to create the next generations of agents.

The parameters of the algorithm and the details of the learning procedure are given in appendix B.\\

{\bf Experiments \& Results}\\
By implementing and executing the method described above, we have been able to train a quantum player for the slime volley game\cite{slimevolleygym}, a simple 2 player game environment with a limited action space size. Making multiple experiments changing the hyperparameters of the evolution, we have been able to beat the default baseline agent implemented in slime volley environment. 

Fig. 6 and 7 shows the evolution of fitness of our two best agents over 1000 generations using parameterized NAND formula evaluation to choose their action as it plays against a classical agent. We can observe that the genetic algorithm first stagnates for about 400 generations. It is possible that better heuristics or method for the neuroevolution\cite{improving_deepneuro} could improve the ability of the model to find a good strategy.

Another RL agent has been trained on the same problem using similar neural network and optimization method but using classical reinforcement learning methods. For now they give performances that are similar to quantum agents but we could hope the quantum agent to achieve better performances by exploring deeper vertices of the game tree with more simulation resources.

\begin{figure}[h]
	\centering
	\begin{minipage}{.45\columnwidth}
		\centering
		\includegraphics[width=\textwidth]{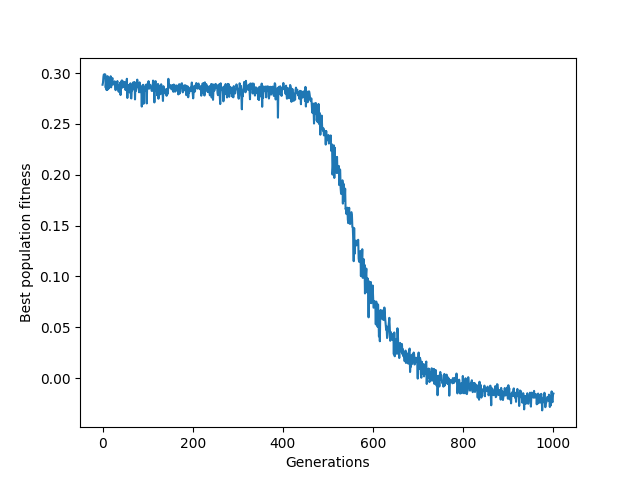}
		\caption{Best Agent}
		\label{agent1}
	\end{minipage}%
	\begin{minipage}{.45\columnwidth}
		\centering
		\includegraphics[width=\textwidth]{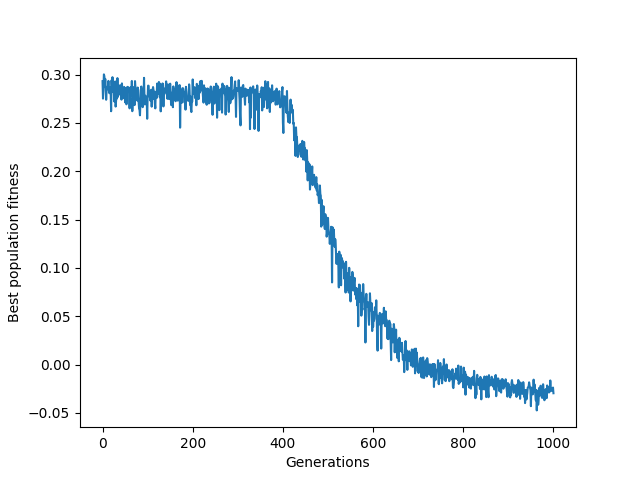}
		\caption{2nd best agent}
		\label{agent2}
	\end{minipage}
\end{figure}

For the experiment presented in this paper, a neural network has been used to complete the capacity of parameterized quantum circuits as we do not dispose of enough qubits or simulation power to deal with enough qubits. A fully-quantum agent could be implemented on a simpler game using a reasonable number of qubits.\\

\textbf{limitations and critics of the learning method}\\
The method used for the training of the agent presents limitations :
As we evaluate fitness of agents on their ability to play a game and so on the quality of the actions they have taken (depending on the result of the NAND formula evaluation), we can't be sure that the oracle proposed by the neural network gives an appropriate knowledge about the leaves of the tree. It could either be a proposition of leaves values that leads to the same NAND evaluation result but does not give an appropriate vision of the game tree.

\section{CONCLUSIONS}

In this article, we have presented a simple method to build circuits for quantum walks on binary trees using basic quantum gates. We have applied this method in the implementation of a quantum algorithm to evaluate boolean NAND formulas using quantum walk on a NAND tree. Finally, we have shown that we could build a quantum agent for a two player game by training a circuit to take the place of the oracle in the NAND formula evaluation algorithm.


\bibliographystyle{IEEEtran}
\bibliography{main} 

\newpage

\onecolumn

\section*{APPENDICES}

\subsection{Full quantum circuit for quantum walk step in NAND formula algorithm (without diffusion step)}
\makebox[\linewidth]{
  \includegraphics[width=1\textwidth]{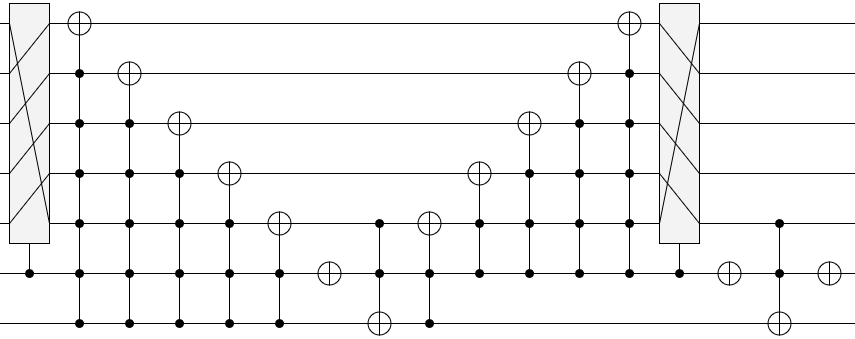}}
\label{walkstep}


\twocolumn


\subsection{Parameters and details on experiments to train the agent}

The way the agent takes decisions for the game is the following :
A simple feed-forward neural network made of 6 layers of 80 neurons takes as input a vector of basic information about the state of the game (position and speed of players and ball) and gives as output a matrix describing which qubits of the \(W\) register will be used to control the $Z$ operator that defines the phase flip of the oracle for each possible move, allowing us to compute the value of each immediately accessible vertices of the game tree. The agents thus chooses randomly one action among all the actions evaluated by NAND computation as winning moves (which keeps us in winning position).

As the simulation time grows with the exploration depth, we have chosen to train our agent with small exploration depth (2-3 steps) but this is sufficient to show the ability of the neural network and the oracle circuit to locally fit to the game.

For these experiments, a population of \(P \in [1152, 9162]\) agents is initialized randomly and evolves with following parameter ranges for the genetic algorithm :\\
- Selection of next generation is simply made by k-bests selection (with $k \in [1; \frac{P}{2}] $)\\
- Mutation is made following a Gaussian distribution with a standard deviation of $[0; 0.1]$\\
- Crossovers are made between randomly selected individuals of the k-bests population. They are made by uniform crossover by windows with window size in [50; 1200]\\
The evolution is made for 1000 generations, each agent is evaluated on at least 25 games where it has to beat an already well trained classical agent.
The parameters ranges used for the training of the quantum agent have been chosen by training a fully classical reinforcement learning agent materialized by a neural network with a structure similar to the one used for the experiment and the same genetic algorithm.

The final score of a game is calculated as follows :\\
 - 1 point if the player wins\\
 - 0 points in the event of a draw\\
 - -1 point if the player looses\\
At each step of the game, a discount factor of $\gamma = 0.99$ is applied to the final score to value the agents that wins faster or repel the defeat further.\\
At the end of their evolution, agents are reevaluated on 1000 games.

Score in the table 1 correspond to their average fitness playing against the default trained agent of the slime volley environment (positive if they win, negative if they lose).\\

\begin{table}[!h]
\caption{Results}
\begin{center}
\begin{tabular}{c c c c c}
Mut. proba. & Select. portion & Std dev. & Cross. 
window & Score\\
\hline
1 & 0.1 & 0.1 & 600 & 0.047\\
1 & 0.1 & 0.1 & 300 & 0.031\\
1 & 0.1 & 0.5 & 300 & -0.25\\
\hline
\end{tabular}
\end{center}
\end{table}

Our application is not differentiable and thus impossible to train with gradient-based reinforcement learning methods. It could be interesting to think about a similar differentiable process allowing us to perform a better optimization.\\

Our limited computation power limits us to explore the game tree two or three steps ahead as the simulation time increases with the exploration depth. It is conceivable that ability to explore deeper vertices of the tree would allow us to achieve better performance in the game.\\

\end{document}